Running Head: QUANTUM APPROACH TO HUMOR

A Layperson Introduction to the Quantum Approach to Humor

Liane Gabora and Samantha Thomson
University of British Columbia

and

Kirsty Kitto
Queensland University of Technology

Correspondence regarding this chapter should be addressed to:

Liane Gabora
Department of Psychology, University of British Columbia
Okanagan Campus, Fipke Centre for Innovative Research
3247 University Way
Kelowna BC, Canada V1V 1V7
Email: liane.gabora@ubc.ca
Tel:604-358-7821 or 250-807-9849

# Introduction

Despite our familiarity with and fondness of humor, until relatively recently very little was known about the underlying psychology of this complex and nuanced phenomenon. Recently, however, cognitive psychologists have begun investigating how people understand humor and why we find certain things funny. This chapter introduces a new cognitive approach to modeling humor that we refer to as the 'quantum approach', which will be explained here in intuitive, non-mathematical terms later (a formal treatment can be found in Gabora & Kitto, 2017). What makes the quantum approach a promising candidate for a theory of humor is that it can be useful for representing states of ambiguity, and it defines states and variables with reference to a context. Contextuality and ambiguity both play a key role in humor, which often hangs on an ambiguous word, phrase, or situation that might not make sense, or even be socially acceptable outside the specific context of the joke. The quantum approach does not attempt to explain all aspects of humor, such as the contagious quality of laughter, or why children tease each other, or why people might find it funny when someone is hit in the face with a pie (and laugh even if they know it will happen in advance); what it aims to do is to mathematically represent the underlying cognitive process of "getting" a joke.

After briefly overviewing the relevant historical antecedents of the quantum approach and other related approaches in cognitive psychology, we present the theoretical basis of our approach, and outline a recent study that provides empirical support for it.

## Situating Our Approach within Humor Research

### Psychological Approaches to Humor

Psychologists have approached the study of humor and joke telling from several directions. Social psychology investigates how humor may influence social interaction; health psychology investigates possible therapeutic advantages of humor; personality psychology looks into what traits are characteristic of funny people, and developmental psychology considers how humor changes throughout the lifespan. Our approach falls under the domain of cognitive psychology, which investigates how we process and respond to information we get from our senses through mental processes such as attention, memory, learning, thinking, problem solving, creativity, and perception. The cognitive approach to humor enables us to not just describe what humor is, but understand how and why something is considered humorous.

### Cognitive Psychology Approaches

To understand how psychologists have made headway in understanding humor we must introduce the concept of a schema. A *schema* is a mental framework that helps us interpret and organize new, incoming information by basing it on these events and beliefs that we have encountered before. Schemas take two forms: they can be either static *frames*, as in a cartoon, or dynamically unfolding *scripts*, as in many jokes. For example, consider the following joke:

"Why did the cookie go to the doctor's office? Because it was feeling crummy!"

It activates a script we hold in our minds about what a visit to the doctor entails (feeling ill, or 'crummy'), as well as a script for what eating a cookie entails (crumbs falling, and so forth).

This joke is an example of a *canned joke*, which is one that is developed prior to being told, and which is self-contained, i.e., its humor does not depend on context or outside information. Many jokes—and particularly canned jokes—have a remarkably consistent structure, consisting of a setup and punchline. The, first part of the joke, the *setup*, establishes the preliminary information and context that will be necessary for the joke to make sense. It often introduces schemas that are seemingly *incongruent*, i.e., that don't appear to go together. For example, in the cookie joke, the setup is the first sentence, which is posed as a question ("Why did the cookie go to the doctor's office?"). The simultaneous activation of scripts about cookies and doctors is mildly confusing because normally these schemas would not go together, and more specifically, a cookie would not normally go to the doctor.

The second part of the joke, the *punchline*, generates humor by introducing an unexpected way of uniting the seemingly incongruent schemas. For example, in the above joke the punchline is the second sentence ("Because it was felling crummy!"). It introduces a word that has two meanings, the word "crummy", which is generally used to indicate that one is feeling mildly ill, i.e., as a synonym for 'lousy', but which could also be used to indicate something that causes crumbs to fall. Thus the punchline provides a pun that unexpectedly unites the schema of what happens when one eats a cookie with the schema of a typical visit to the doctor.

To understand a joke's humor the listener must simultaneously hold both the setup and the punchline in their mind (Attardo, 1994; Raskin, 1985). This process of holding seemingly contradictory schemas in one's mind at once is called *bisociation* (Koestler, 1964). The incongruity of bisociation, the part that creates the humor, is a result of the mismatching or contrasting of the schemas within the joke. While in the cookie joke two seemingly incompatible schemas were introduced in the setup, it is often the case that the setup presents only one schema and the punchline a second that is simultaneously incongruous yet somehow compatible with the first schema. The incongruity creates a violation of expectations with respect to how a schema typically unfolds, causing a feeling of surprise. The listener attempts to resolve, or find a logical way of reconciling, the incongruity, and feels relief when the punchline provides this. Because so many jokes involve incongruity resolution, early cognitive theories of humor suggested that the act of resolving incongruity between schemas is always the sole source of a joke's humor (Shultz, 1974; Suls, 1972). However, there is increasing evidence that incongruity resolution is not required for humor, and that even when present it is not the source of humor; instead, the humor arises because the cognitive effort involved in bisociation accentuates the contrast between the different schemas (e.g., Martin, 2007; McGraw and Warren, 2010).

To better understand how the incongruity of schemas and bisociation function, consider the following joke:

"Time flies like an arrow. Fruit flies like a banana."

There are two schemas at work in this joke. One can be found in the setup, where "flies" is interpreted as a verb, and in this context means "goes by very quickly". The second schema is found in the punchline. Here, "flies" is interpreted as a noun and refers to an insect. The incongruity created in this joke is caused by the ambiguity, or double meaning, of the phrase "flies" produced by these contrasting schemas. The second schema—the punchline—violates the expectations established in the first schema—the setup—about how the ambiguous word 'flies' is to be understood. The 'cognitive work' required to understand this joke involves realizing that "flies" has unexpected transitioned from a verb to a noun, thereby uniting or 'bisociating' the seemingly incongruent schemas. The result is an experience of humour potentially accompanied by a mild feeling of relief.

One aim of those who take a cognitive approach to humor is to better understand the conditions that make incongruity funny. To investigate this, Mihalcea, Strapparava, and Pulman (2010) developed an incongruity detection task that involved giving participants a joke setup and then asking them to choose one of the provided multiple choice answers as the punchline. For example, one such item is the following:

"Don't drink and drive. You might hit a bump and"

a) spill your drink.
b) get a flat tire.
c) have an accident.
d) hit your head.

The correct punchline (i.e., the one that was used in the actual joke) was 'a'. In a series of questions like this they systematically varied aspects of the setup and punchline, including the degree of relatedness between them. They found that, of all the punchline candidates, the words in the correct punchline, had the lowest level of relatedness to the words in the setup, and that the presence of features such as alliteration or polysemy, enhanced incongruity in jokes.

**Computational Approaches to Humor**

There are computational models of humour detection and understanding (e.g., Reyes, Rosso, & Veale, 2013), that update the interpretation of an ambiguous word or phrase as new contextual information is parsed. For example, in the "time flies" joke, a computational model would shift from interpreting FLIES as a verb to interpreting it as a noun. There are also computational models of humor that generate jokes through lexical replacement; for example, by replacing a taboo word with a similar-sounding innocent word (e.g., Binsted, Pain, & Ritchie, 1997; Valitutti, Toivonen, Doucet, & Toivanen, 2013). Perhaps the most well-known computational models of humor is the Joke Analysis and Production Engine (JAPE), developed by Ritchie (2001). JAPE creates puns and riddles that consist of a setup and punchline and that contain incongruity. An example of a joke generated by JAPE is:

"What do you call a quirky quantifier? An odd number."

Although computational models of humor are an exciting development, such models are still limited with respect to the type of jokes they can generate, i.e., they are generally limited to simple wordplay such as puns. Furthermore, the existence of such models does not necessarily tell us when jokes are humorous, or why.

These computational approaches to humor are provocative, and occasionally generate laugh-worthy jokes. However, while they might tell us something about the generation of verbal humor we claim that they do not provide an accurate model of the cognitive state of a human mind at the instant of perceiving a joke. As mentioned above, psychologists believe that humor often involves not just shifting from one interpretation of an ambiguous element to another, but the simultaneous holding in mind of the interpretation that was perceived to be relevant during

5
QUANTUM APPROACH TO HUMORthe set-up and the interpretation that is perceived to be relevant during the punchline. We turned to the generalized quantum formalism explicitly because it enables us to model the cognitive state of holding two schemas in mind simultaneously.

**Quantum Theory**

One way to conceptualize humor and the above ideas in a cognitive approach is by using Quantum Theory, a formal framework first used in quantum mechanics that has since been shown useful for modeling cognitive states that involve context-dependency and ambiguity. In recent years, there has been an explosion of research applying Quantum Theory to psychological phenomena, in areas that include information retrieval (Van Rijsbergen, 2004; Melucci, 2008), combinations of words and concepts (Aerts, 2009; Aerts & Gabora, 2005a,b; Bruza et al., 2009, 2015; Gabora, 2001; Gabora & Aerts, 2002, 2009), as well as social science (Haven & Khrennikov, 2013; Kitto & Boschetti, 2013) and creativity (Gabora & Aerts, 2009; Gabora & Kitto, 2013; Gabora & Carbert, 2015). As mentioned in the introduction, it is the capacity of this theory to formally represent states of context-dependency and ambiguity that motivate its application in humor research.

Consider an ambiguous word, phrase, or situation (such as the word "flies" in the joke above) that has a hidden or unexpected interpretation (as an insect rather than present tense of the verb "to fly").-Quantum models are useful for describing situations involving this type of *potentiality*, where unknown contextual factors can change the way in which a concept is interpreted. This property of quantum models has been used to good effect in modeling the cultural analog of exaptation, wherein an idea that was originally developed to solve one problem is applied to a different problem (Gabora et al., 2013). For example, consider the invention of the tire swing. It came into existence when someone reconceived the tire, originally a functioning part of a car, as an object that could form part of a swing that one sits on. This repurposing of an object designed for one use for use in another context is referred to *cultural exaptation*. We can easily see how cultural exaptation parallels incongruity rather well by considering the above "time flies" joke once more. The way in which the ambiguous word, "flies", first functions as a verb in the setup, and then as a noun in the different context presented by the punchline is similar to how the tire's use changes from when it is part of a car to when it is used as a swing. It is from this ambiguity that quantum theory supposes cognitive humor arises from. It is this kind of ambiguity and context-driven change of state that quantum theory was expressly developed for.

The Quantum Theory of Humor (QTH) described in this chapter builds upon a preexisting model called the *State-Context-Property (SCOP) Model* (Gabora & Aerts, 2002). The model consists of three main variables (as reflected in its title). The first is the *State Space*. When the SCOP model is applied to humor, this includes all the possible interpretations an ambiguous element can have, and the activated schemas within all of these interpretations. The second variable is the *Context*, which changes as the setup unfolds and new information becomes available. Additionally, the interpretation of a joke is contextually influenced by the teller of the joke, the surroundings of when it is told, and the prevailing mood and atmosphere. The third variable brings the previous two together and it is the *Transition Probability*, which gives the probability of changing from one state to another under the influence of a particular context. In other words, this third variable raises the following question: assuming that the interpretation, or the 'getting' of the joke, is dependent on the contextual factors mentioned above, what is the likelihood that the receiver of a joke will interpret it in a particular way?

The QTH attempts to model the process of 'getting' a joke by considering the way in which a person adjusts their interpretation to be consistent with the unfolding contextual information that becomes available as the joke is told. Let us consider the "fruit flies" pun from the last section one more time. When the setup of the joke ("time flies like an arrow") is read as the only context, 'flies' is a verb. When the context of the punchline ("fruit flies like a banana") is revealed, 'flies' is more naturally interpreted as a noun, which carries a different meaning and therefore adds incongruity to the original interpretation. This changes the mental state of the listener to one that entails ambiguity. This ambiguity is represented as a superposition state (i.e., a vector) in a high dimensional semantic space representing the possible interpretations available for the listener to give to the joke. As the listener hears the joke, more context is provided and the understanding (or the cognitive state of the listener) unfolds according to the transition probabilities associated with the cognitive state and the context. At the point where the listener interprets the joke, the cognitive state changes to a state of perceiving a bisociation, through a process we model as 'collapse to a superposition state'. Since funniness is yoked to bisociation, the listener now finds the joke funny, or does not. The probability associated with the listener finding the joke funny is estimated by considering the projection of the interpretation onto a set of orthogonal axes (or 'basis states'), which are embedded in the higher dimensional semantic space that represents the interpretation that a listener is attributing to the joke. This projection is used to represent extreme states of funniness and absolute non-funniness.

Testing whether this intuitive model of humor is correct would require experimental data. We know that cognitive systems which violate the *law of total probability* (LTP) are often well modeled by the quantum formalism. The LTP states that the probability of some observable event should satisfy the distributive axiom, i.e., the total probability of the observable event should be equal to the sum of the probabilities of it under the possible sets of more specific conditions. Thus, if jokes satisfy LTP then the probability of the joke being judged funny should be equal to the sum of the probability of it being judged funny *given different possible semantic interpretations*. It was hypothesized that since the funniness of a joke generally relies upon one specific semantic framing or context, this statement was very likely to be false.

## An Empirical Test of the Theory

**Methods**

A recent study was carried out to test the predictions of the quantum theory of humor at the Okanagan campus of the University of British Columbia (Gabora & Kitto, 2016). A class of twenty-one undergraduate university students enrolled in a 'Psychology of Humor' course were presented with a questionnaire consisting of a series of jokes. The study was explained and carried out by an undergraduate research assistant (the second author of this paper). The professor of the course (the first author) was not present at the time when the research project was explained, nor when the study was carried out, so as to make the students feel they were not obligated to participate. To also ensure their free and informed consent, they were told that at any point in time they were free to discontinue the study without any negative consequences. To further ensure this, they were also told that the professor would not have any knowledge of who participated in the study, would not have access to the raw data until the course was completely finished, and that participation in the study would not affect their standing in the course of with the University. The participants were told that the purpose of the study was to contribute to a broader understanding of the cognitive processes involved in the 'getting' of a joke and that it would take approximately 20 minutes to complete. Before being given the questionnaire, they signed consent forms to ensure that they understood their participation was voluntary. They were then asked to rate each joke on a 5-point Likert scale from one (1) for jokes they consider 'Not Funny At All' to five (5) for jokes they consider 'Hilarious'.

**The Stimuli**

The questionnaire consisted of randomly ordered jokes and joke variants that came from seven joke sets. A joke set started with an original and complete canned joke which, as stated earlier in this chapter, is a joke that is developed prior to being told, and which is self-contained. The complete canned joke was then taken and divided into joke fragments consisting of either the setup or punchline isolated from the rest of the joke.

To illustrate how variants were created, consider the following original joke:

"Why was 6 afraid of 7? Because 7 ate 9!"

To create joke fragments, we first must identify the components of this joke which are the setup and punchline. The setup is the part of the joke providing preliminary information, most often found at the beginning of the joke, therefore "Why was 6 afraid of 7?" is the setup. The punchline is the second part, often containing the resolution to the joke's incongruity as well as the humor, so it is "Because 7 ate 9!". For each joke, both the setup and punchline were presented isolated, as shown above, to the participants, and then rated for funniness just as the original canned joke was.

A joke set also consisted of variants created from the original jokes that were made with either incongruent or congruent schemas. As mentioned earlier in this chapter, incongruent schemas those that suggest different interpretations or contexts for an ambiguous joke element. This type of variant was expected to be comparable in funniness to the original because of the presence of incongruency and resolution. Congruent schemas did not introduce new or ambiguous information in the setup or punchline. Congruence was used in this study to extend the understanding of the way in which different contexts could affect the perceived funniness of a given joke.

The sample joke above uses a pun of the words 'ate' and 'eight' to create humor. To produce an incongruent variant, the ambiguity must be maintained in the joke but the setup or punchline must change with respect to content. By creating incongruent variants we were able to determine to what extent it was the incongruity itself that created the humor of the original joke versus to what extent it was the content of the joke. These variants were generated by either altering the setup and keeping the punchline the same, or by altering the punchline and keeping the setup the same. An incongruent variant of the above joke with the same setup but an altered punchline is:

"Why was 6 afraid of 7? Because 7 was a registered six offender!"

This variant maintains the humor and unexpected nature and incongruity of the punchline of the original joke, by creating a pun of 'sex' out of the word 'six', but contains different content and allows for a different experience and surprise by the listener. Now, to illustrate how congruent variants created consider the below joke that we are revisiting from earlier in this chapter:

"Time flies like an arrow; fruit flies like a bird."

Congruence can be achieved by modifying the setup to make it congruent with the punchline, or by modifying the punchline to make it congruent with the setup. To create a congruent variant, the ambiguity between the setup and punchline must be removed. Often, once the ambiguity between schemas is removed the humor is also removed because the joke appears to simply be stated fact or normal dialogue which a reader would not find humorous. For example, if the setup makes use of a noun, like our sample joke above does, then a congruent modification of the punchline would also use a noun. A congruent variant of the above joke with an altered punchline is:

"Time flies like an arrow; time flies like a bird."

The congruent variant of this punchline does not give the receiver of the joke much sense of surprise because it is not in contrast to the schema that was introduced in the setup. As a result, there is considerably less humor to be found in that variant. It is also possible alter the setup to be congruent with the information in the punchline:
"Horses like carrots; fruit flies like a banana."
Now that we have seen how the joke variants were created for this study, we provide an example of an entire joke set in Table 2. Seven such sets were put into randomized order and rated in funniness by participants.

| Joke Variant | Example of Joke Variant |
| --- | --- |
| Original | "Time flies like an arrow; fruit flies like a banana." |
| Set-up Only | "Time flies like an arrow." |
| Punchline Only | "Fruit flies like a banana." |
| Congruent Set-up | "Horses like carrots; fruit flies like a banana." |
| Congruent Punchline | Time flies like an arrow; time flies like a bird." |
| Incongruent Set-up | "Time flies like a bird; fruit flies like a banana." |
| Incongruent Punchline | "Time flies like an arrow; fruit flies like an apple." |

*Table 1:* Different kinds of jokes and joke fragments used in the study (for details see text).

**Results**

To determine how funny the jokes were rated in relation to each other, we calculated the average funniness rating for each joke type and variant. These data are summarized in Table 2 As expected, the participants rated original jokes with the highest funniness ($M=2.96$), incongruent variants (incongruent punchline $M=2.58$, incongruent setup=2.69) as second funniest. When the incongruity was absent, either in the form of setup fragments ($M=1.23$), punchline fragments ($M=1.13$), or a joke variant that was altered to remove incongruity (congruent punchline $M=1.53$, congruent setup $M=1.83$), they were all rated as significantly less funny. Occasionally, a joke variant that retained the incongruity was rated as even funnier than the original, as was the case with the incongruent variant mentioned above ("Why was 6 afraid of 7? Because 7 was a registered six offender!"). This is perhaps due to the participants being familiar with the original joke and therefore finding the new incongruity more unexpected and bizarre, resulting in overall higher funniness rating. The fact that an incongruent variant was found to be funnier than the original supported the hypothesis that incongruence plays an important role in humor.

| Joke Variant | Mean Funniness Rating |
|---|---|
| Original | 2.96 |
| Set-up Only | 1.23 |
| Punchline Only | 1.13 |
| Congruent Set-up | 1.83 |
| Congruent Punchline | 1.53 |
| Incongruent Set-up | 2.69 |
| Incongruent Punchline | 2.58 |

*Table 2:* The mean funniness ratings of each kind of joke variant across all participants of each joke set.

This study provides support for the Quantum Theory of Humor (QTH) because the data obtained appears to violate the Law of Total Probability (LTP) for some jokes, as was predicted.

**Future Work**

This study was the first to consider applying quantum modeling to humor research. However, this study was conducted with a rather small sample size. Another limitation is that the inclusion of cultural references in some of the jokes may bring about confusion on the part of participants who are unfamiliar with them. For example, consider the following joke, which was used in the study:
 "I am sick of having to go to two different huts for pizza and sunglasses."
To understand this joke, the person reading it would need to know that there are two stores called Pizza Hut and Sunglasses Hut. The understanding of the joke therefore depends on contextual information that may or may not have been obtained prior to hearing the joke, depending on where the participant grew up. Finally, the construction of incongruent and congruent joke variants is currently an art rather than a science, and more work will be required to explore this issue. In particular, it would be useful to construct a systematic study of the manner in which adjusting the congruence of the setups and punchlines influence the subject's perception of the joke, and to make use of the geometric aspect of quantum theory to explain this effect.

## Conclusion

This chapter provided a laypersons introduction to a new cognitive approach to the study of humor that uses quantum theory to model bisociation and incongruity resolution. We hope that it can help elucidate the cognitive processes involved in 'getting' a joke, and shed light on how a successful joke can be structured. We caution that, despite the intuitive appeal of the approach, it is still rudimentary, and more research is needed to determine to what extent it is consistent with empirical data. Nevertheless, we believe the fledgling research program outlined in this chapter promises to be an exciting step toward a formal theory of humor. A follow-up study is currently underway with a larger and more diverse sample as well as an expanded version of the questionnaire containing more joke sets. Thus, further research is building upon the modest foundation described here. In addition, we hope that this work sheds light on peoples' own everyday experience of humor. Perhaps next time you laugh at a joke you will find yourself analyzing it from a cognitive perspective. For example, you may ask yourself: how did the structure of the joke influence your comprehension of it? Can you identify if and how incongruity was involved? How would you explain the joke using the quantum theory of humor? Although there is much to be discovered about the elusive phenomenon of humor, the capacity to formally model the process of getting a joke is an exciting step forward toward understanding the cognitive processes involved in joke telling and comprehension.

## Acknowledgements

This work was supported by a grant (62R06523) from the Natural Sciences and Engineering Research Council of Canada.